# Hot Dirac Fermions in Epitaxial Graphene

Dong Sun[1], Zong-Kwei Wu[1], Charles Divin[1], Xuebin Li[2], Claire Berger[2], Walt A. de Heer[2], Phillip N. First[2], and Theodore B. Norris[1]

[1]Center for Ultrafast Optical Science, University of Michigan,

Ann Arbor, MI 48109-2099

[2]School of Physics, Georgia Institute of Technology, Atlanta, GA 30332

**Abstract**

We investigate the ultrafast relaxation dynamics of hot Dirac fermionic quasiparticles in multilayer epitaxial graphene using ultrafast optical differential transmission spectroscopy. We observe DT spectra which are well described by interband transitions with no electron-hole interaction. Following the initial thermalization and emission of high-energy phonons, the electron cooling is determined by electron-acoustic phonon scattering, found to occur on the time scale of 1 ps for highly doped layers, and 4-11 ps in undoped layers. The spectra also provide strong evidence for the multilayer stucture and doping profile of thermally grown epitaxial graphene on SiC.

Graphene is a planar layer of carbon atoms in a hexagonal lattice, with a linear energy spectrum near the intersection of the electron and hole cones in the band structure (the Dirac point)[1-4]. The linear energy dispersion gives rise to massless Dirac quasiparticles with an energy-independent velocity, leading to unusual quantum properties[5-9]. and a predicted electrodynamic response differing significantly from that of fermionic plasmas with parabolic dispersion[10, 11]. Graphene grown expitaxially on SiC substrates has been proposed as a platform for carbon-based nanoelectronics[12, 13]. While transport in steady-state measurements is controlled by the electrons near the Fermi level, transport in high speed devices is determined by the dynamic conductivity of hot carriers, *i.e.* electrons whose temperature is elevated above the lattice temperature due to the presence of high fields and/or dynamic fields in the material. It is critical therefore to understand both the cooling of the electrons due to coupling to lattice phonons, and to understand the response of the quasi-particle plasma to dynamical fields.

Here we report the first ultrafast spectroscopy of hot Dirac quasiparticles in graphene in the region near the Fermi level[14]. As illustrated in figure 1, a 100-fs near-infrared (800-nm) optical pulse excites quasiparticles from the valence to the conduction band across the Dirac point; the optical response of a multilayer graphene structure containing both doped and undoped layers is measured via the differential transmission (DT) of a tunable probe pulse as a function of pump-probe delay. The electrons have an initial energy of 428 meV above the Fermi level in the doped layers and 777 meV in the undoped layers. Due to rapid carrier-carrier scattering, a hot thermal distribution is established within the time resolution of the experiment (150 fs). This distribution cools towards the lattice temperature, initially via the emission of high-energy (194 meV and 330 meV) phonons[15], and later via the interaction with acoustic phonons. In these experiments, the elevated temperature of the quasiparticles is manifested primarily through the modification of the probe absorption by Pauli blocking of interband transitions.

For the optical experiments, a 100-fs 250-kHz amplified Ti: Sapphire laser at 800 nm pumps an infrared optical parametric amplifier (OPA) with signal wavelength tunable from 1.1 to 1.6 µm and idler wavelength tunable from 1.6-2.6 µm. The 150-fs idler is used as the probe and the 800-nm beam is the pump; the beams are co-linearly polarized and focused to 40-µm and 80-µm diameter spots respectively on the sample. The probe beam after the sample is filtered in a monochromator with 3-nm resolution and detected by an InGaAs photo detector and lock-in amplifier referenced to the 4.2-kHz mechanically chopped pump. The sample temperature is controlled over a range 10-300 K.

The sample is an ultrathin epitaxial graphene film produced on the C-terminated ($000\bar{1}$) face of single-crystal 4H-SiC by thermal desorption of Si[16]. Figure 1 shows the structure of the sample[13,16,17]: the first carbon layer (green) is covalently bonded to the 4H-SiC substrate and acts as a buffer layer; the following layer (red) exhibits the graphene electronic spectrum and is doped by charges transferred from the SiC. From the measured Fermi level (see below), the charge density is estimated to be $9 \times 10^{12}$ electrons/cm$^2$. The graphene layers (blue) above the doped layer are essentially neutral[17,18]. For the growth conditions employed, the number of neutral layers has been estimated to be in the range 15-20.[19,20].

Figure 2a shows representative DT spectra at various probe time delays. The DT amplitude peaks near zero time delay for all probe wavelengths, consistent with the establishment of a hot thermal carrier distribution within the experimental time resolution. The DT amplitude then relaxes towards zero on a time scale of 15 ps. The DT signal flips from positive on the blue (high-energy) side of a probe wavelength of 1.78 μm to negative on the red (low-energy) side of 1.78 μm, and flips back from negative to positive again at 2.35 μm.

Figures 2 (b) and (c) show DT time scans for selected probe wavelengths on both red and blue sides of the two zero crossings. Immediately following the pump pulse, the DT signal is positive over the entire probe spectral range. The DT signal becomes negative within 2 ps if the probe wavelength falls between 1.78 μm and 2.35 μm, otherwise it remains positive until the signal decays away. The DT signal relaxes to zero on the time scale of 1-10 ps depending on probe wavelength. The lack of the data in the two blank regions of Fig. 2(a) is due to limitations in tuning our OPA.

Fig. 3 shows the effect of lattice temperature on the carrier dynamics. For spectral regions where the DT sign is positive, the DT dynamics show little temperature dependence apart from minor amplitude changes and slightly different relaxation times. When the probe wavelength falls between 1.78 µm and 2.35 µm, the DT signal is positive at early times, and becomes negative within a few picoseconds. The amplitude of the negative DT component decreases with increasing temperature and almost disappears for temperatures above 180 K. The delay time at which the DT crosses zero increases monotonically with temperature (Figure 3a inset).

We now turn to the interpretation of the DT and the origin of the zero-crossings. From the simplest point of view, the differential probe transmission spectrum arises from the change in carrier occupation functions in the bands, since generally the probe absorption is proportional to $f_v(1-f_c)$ where $f_v$ ($f_c$) is the occupation probability in the valence (conduction) band. Following the excitation of quasiparticles high into the conduction band by the pump, electron-electron scattering on a time scale short compared to 150 fs establishes a hot thermal distribution. Since the carrier occupation probability above the Fermi energy is increased by the pump, the DT signal is positive due to reduced probe absorption. Below the Fermi level, the heating of the electron plasma reduces the occupation probability, so the DT is negative. Thus the upper zero crossing at 1.78 µm probe wavelength arises from the smearing of the Fermi level in the doped layers. We find the Fermi level $\mu_d$ for the doped layer to be 348 meV above the Dirac point. This result is close to the calculated Fermi level in the carbon-deficient geometry[17], and is consistent with the results of transport studies on epitaxial graphene grown on the C-terminated ($000\bar{1}$) face[13]. We note additionally that there is no peak in the DT spectrum near the Fermi level; this indicates that there is no Fermi edge singularity[21, 22] due to electron-hole interactions in the interband absorption spectrum of graphene, as may be expected from the massless nature of the quasiparticles.

At very long probe wavelengths, i.e. for transition final states well below the Fermi level of the doped layers, one may expect the DT spectra to be determined primarily by the carrier occupations in the undoped layers; since the pump pulse generates hot carriers in the undoped layers, the sign of the DT signal arising from the undoped layers should be positive for all wavelengths. However, for probe wavelengths below the Fermi level of the doped layer, the contribution of the doped layer to the DT is negative. Thus one expects that for some probe energy the net DT signal should flip sign; this is the origin of the lower zero crossing at 2.35 µm.

The probe transmission and reflection spectra were modeled using the transfer-matrix method[23], incorporating both the interband and intraband contributions to the dynamic conductivity (or dielectric constant) of each graphene layer.[4, 11] We calculated the transmission spectrum of the multilayer structure of figure 1, with transfer matrices for 20 undoped graphene layers (Fermi level at the Dirac point) and 1 doped layer[21] with a Fermi level of 350 meV.

Figure 4 shows the calculated transmission spectrum for various electron temperatures; where for simplicity we have assumed the temperature to be the same for all layers. At low temperature (10K), the transmission spectrum shows an absorption edge at $\hbar\omega = 2\mu_d$, as one would expect from the simple picture of interband absorption discussed previously. As the temperature increases, the absorption edge due to the doped layer broadens due to smearing of the carrier distribution around the Fermi level, and the undoped layers contribute a broad peak at low energy.

Figure 4b shows the calculated DT spectrum for an initial electron temperature of 10 K (i.e. the DT spectrum is the transmission spectrum for an elevated electron temperature minus the transmission spectrum for 10 K). The DT spectra show the upper and lower zero crossings at energies ($\hbar\omega = 2\pi_d$ and $\hbar\omega \approx 1.5\pi_d$ respectively) close to those observed in the experiment. Immediately following the pump pulse, the initial hot electron temperature is higher than 1200 K for our experimental excitation intensity; from the simulation this implies a positive DT signal over the entire spectral range, exactly as observed.

Examination of Figure 4 shows that when the probe energy is between the two DT zero crossings, the transmission curve does not relax monotonically with decreasing electron temperature; it decreases to a minimum around 400 K and then turns back and increases with decreasing temperature. The high initial electron temperature gives a positive DT signal; as the carriers lose energy, the DT amplitude decreases and becomes zero for an electron temperature of approximately 700 K. The DT is then negative and reaches its maximum negative amplitude for an electron temperature of 400 K. With further cooling, the DT approaches zero. The time delay at which the DT flips sign should be expected from the model to increase with lattice temperature, as observed in the experiment (inset of Fig. 3a).

In contrast, when the probe wavelength is either above the upper or below the lower zero crossing, the transmission decreases monotonically with the electron temperature and the DT decay curves are only weakly dependent on lattice temperature, as is apparent in Figs. 3b and 3c.

Additional simulations performed by excluding various contributions to the total conductivity reveal that the dominant contribution is the real part of the interband conductivity. We note that the observed lower zero crossing is unrelated to the TE mode arising from the negative imaginary part of the interband conductivity predicted in recent work[11]. The imaginary intraband conductivity only results in a small shift of the lower zero crossing (less than 5%), and this contribution cannot be isolated in our normal-incidence DT experiment. Our DT spectra are well described by interband transitions and the single-particle density of states for linear dispersion, and no electron-hole interaction. We also note that the simulations are sensitive to the detailed layer structure of the epitaxial graphene, and the spectra provide strong corroborating evidence for the epitaxial graphene structure determined using other methods[16, 20, 24-26]; our best fit to the data is obtained for one conducting layer and 20 undoped layers.

Our calculations of the DT assume that the quasiparticle plasma can be described by a thermal distribution characterized by a single electron temperature $T_e$ in all layers. Within the time resolution of our experiment, the electron distribution appears to be thermal in all experimental DT spectra. From the time delay of zero crossing DT point at 10 K, we find that the hot electron temperature relaxes to around $T_e = 420$ K (36 meV) on the time scale of 1.7 ps by emitting two or three 197 meV[15] optical G phonons[27, 28], or one to two 330 meV D phonons[15]. The relaxation afterwards is mainly due to the relatively slow acoustic phonon scattering process.

To determine the electron temperature $T_e$ as a function of time from the DT amplitudes, we calculate the DT at a given probe wavelength for each value of the temperature. We find that fits of the DT dynamics near the upper zero crossing (where the DT is dominated by the doped layer) show that the decay of $T_e$ is reasonably well described by a single exponential with a time constant of 1.45 ps. Fits to the DT dynamics near the lower zero crossing can not be fit with single exponential decay of $T_e$, but can be well fit by a stretched exponential $\exp\left[-(t/t)^{1/h}\right]$ with τ in the range of 4-11 ps and a heterogeneity parameter $h=3$. The long nonexponential decay may arise from the generation of hot phonons [27, 28] although disorder in the sample or the temperature- and density-dependence of the scattering rate may also contribute. The observed time scale for the cooling is consistent with the 4-ps electron-phonon scattering time constant estimated recently from magnetoresistance measurements[13] and with recent calculations[27].

### Acknowledgements

This project has been supported by NSF grants ECCS-0404084 and ECCS-0521041.

Correspondence and requests for materials should be addressed to tnorris@eecs.umich.edu

**Figure Captions:**

**FIG. 1 Sample structure and energy dispersion curve.** Sample structure and energy dispersion curves of doped and undoped graphene layers. The sample has a buffer layer (green) on the SiC substrate followed by 1 heavily doped layer (red) and approximately 20 undoped layers (blue) on top. The Fermi level is labeled with a dashed line (brown) lying 348 meV (from the later data) above the Dirac point of the doped graphene layer and passing through the Dirac point of the undoped graphene layers. The blue solid line shows the transitions induced by the 800-nm optical pump pulse; the three dashed lines correspond to probe transitions at different energies with respect to the Fermi level (discussed in the text).

**FIG. 2 DT spectrum and zero crossings. a,** DT spectrum on epitaxial graphene at 10K, with 500-µW 800-nm pump (photon fluence of $1.6 \times 10^{14}$ photons/cm$^2$ per pulse), at probe delays of 10 ps, 5 ps, 2 ps, 1 ps, 0.5 ps, and background (50 ps before the pump arrives). The arrows at 1.78µm and 2.35µm indicate the DT zero crossings. **b,** DT time scan of the two probe wavelengths marked in part a at the red (1.85µm) and blue side (1.75µm) of the 1.78µm DT zero crossing. **c,** Time scan of the two probe wavelengths marked in part (a) at the red (2.40µm) and blue side (2.25µm) of the 2.35µm DT zero crossing. In all figures, the dashed line (brown) marks where the DT signal is zero. The DT tails in b and c are simply fitted by a sigmoidal curve.

**FIG. 3 Temperature dependent DT spectrum. a,** DT time scans at temperature 10K, 30K, 50K, 77K, 130K and 180K with 500µW pump at 800nm and 2.25µm probe. The DT time scans were fit with a sigmoidal curve to show more clearly the behavior of the zero crossings. In the inset the DT zero-crossing points at different temperatures are marked with different colors. **b,** DT time scans at temperatures of 10K, 77K and 287K with 1mW pump at 800nm and 1.57µm probe. **c,** DT time scans at temperatures of 77K and 290K with 1mW, 800nm pump and 2.4µm probe. In all figures, the dashed line (brown) marks where the DT is zero.

**FIG. 4 DT signal simulation. a,** Simulated transmission curves at different electron temperatures. In the inset, the transmission curves at low electron temperature are shown expanded for frequencies around the two DT zero-crossings. **b,** Simulated DT/T curves at different electron temperatures with lattice temperature at 10K. In the inset, the DT/T curves for low electron temperatures are expanded in the vicinity of the two DT zero crossings. Both figures share the same legend.

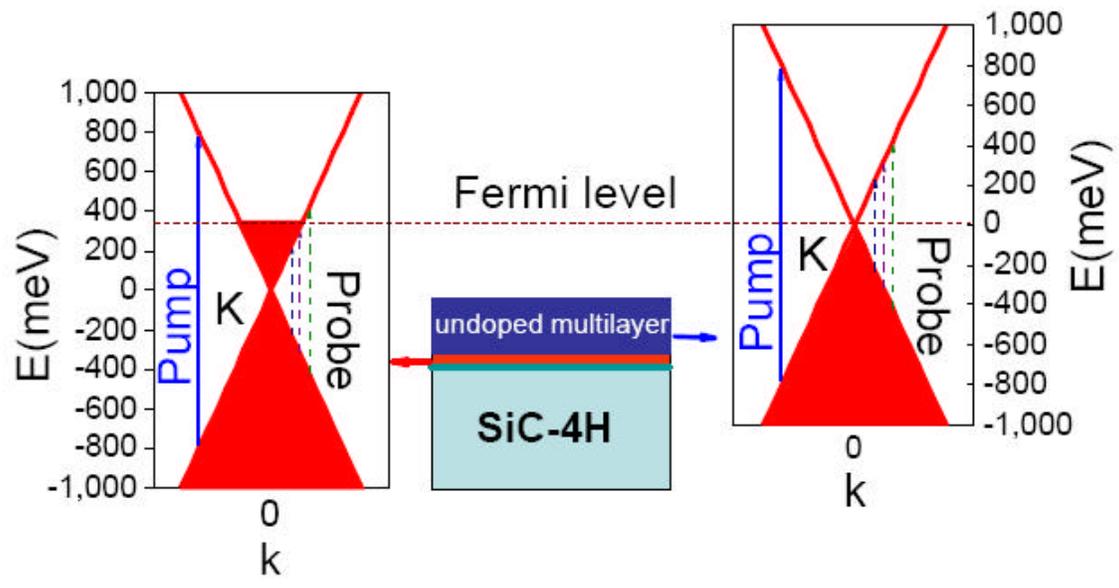

Fig. 1.

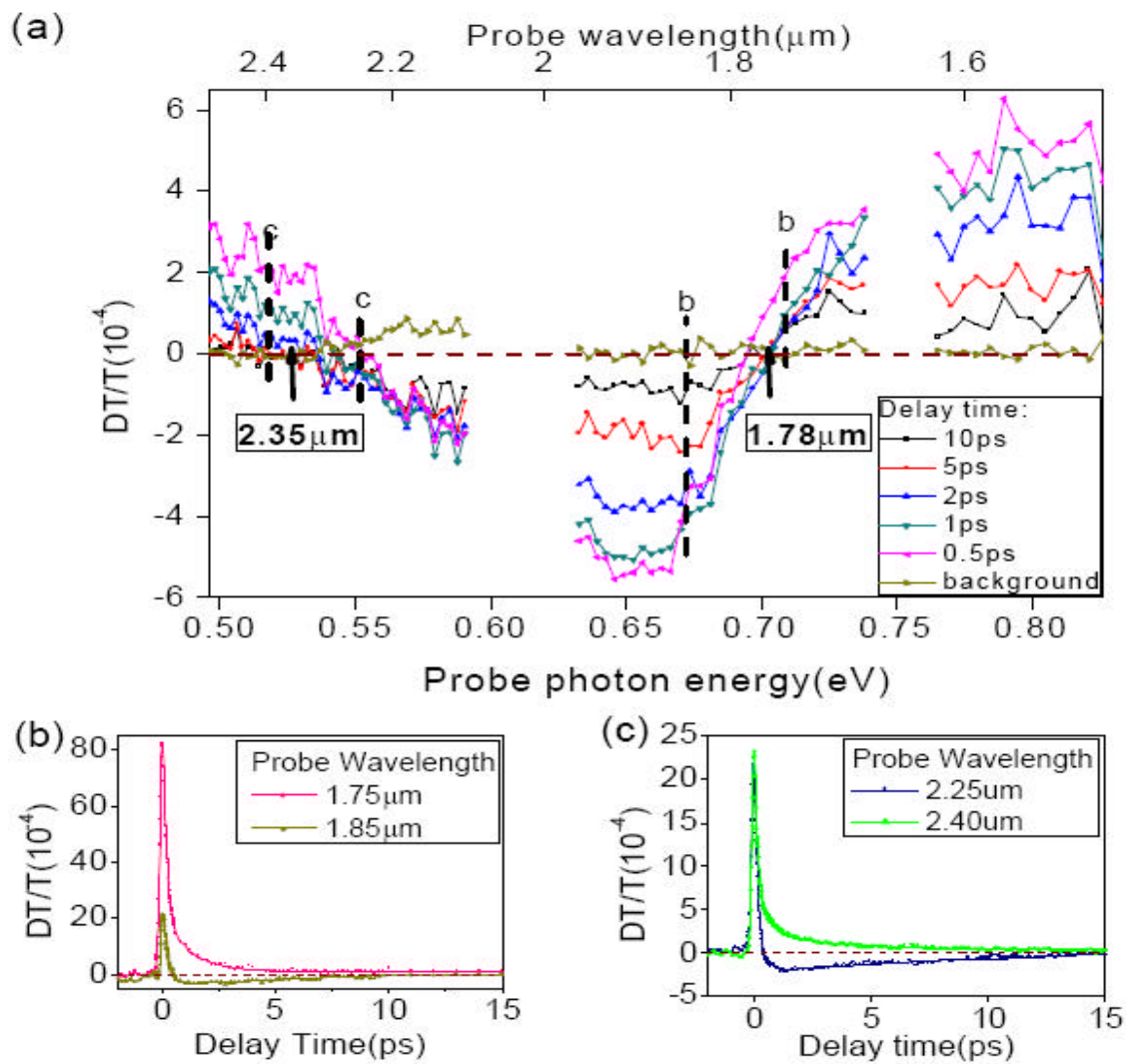

Fig. 2

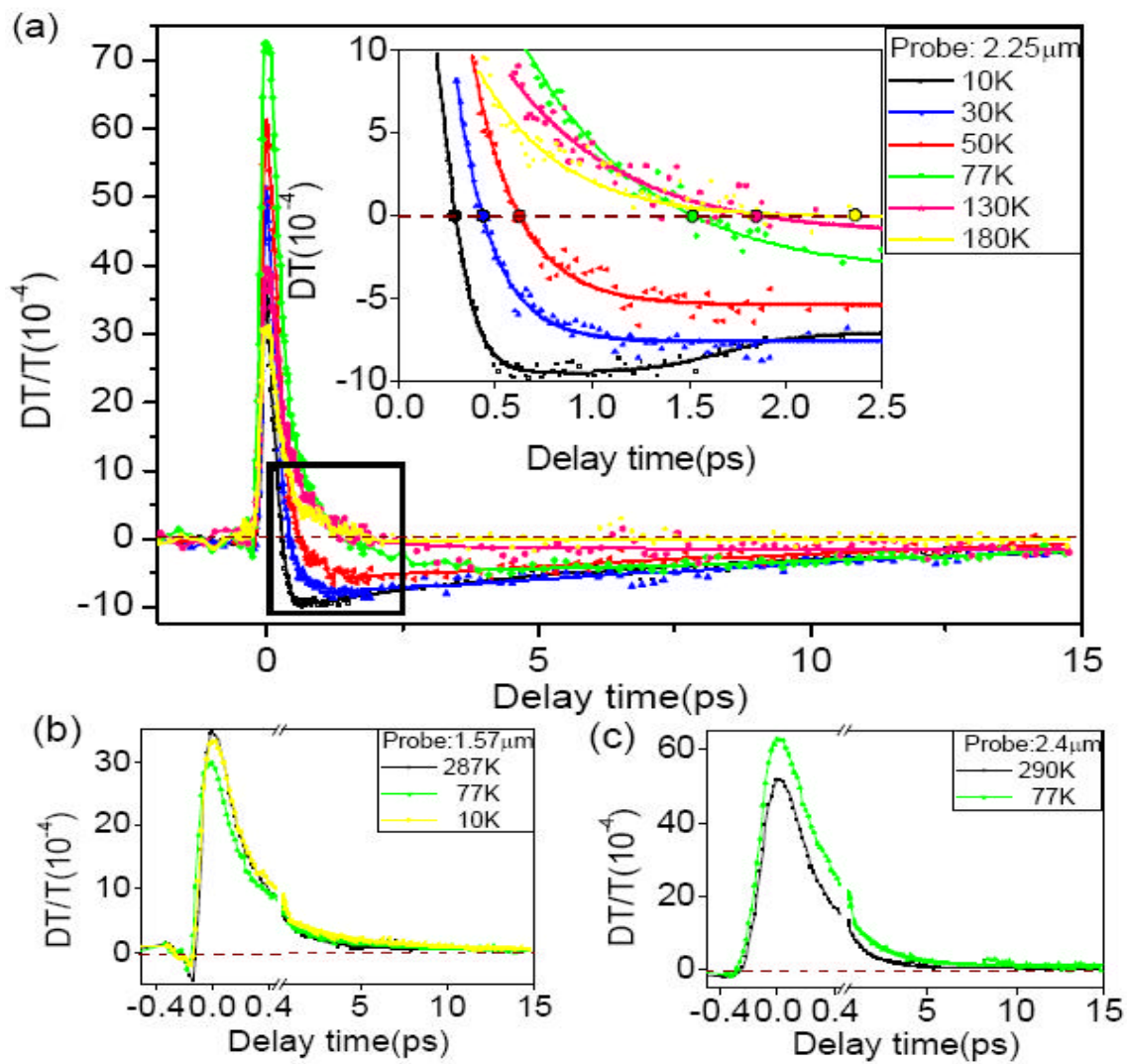

Fig. 3

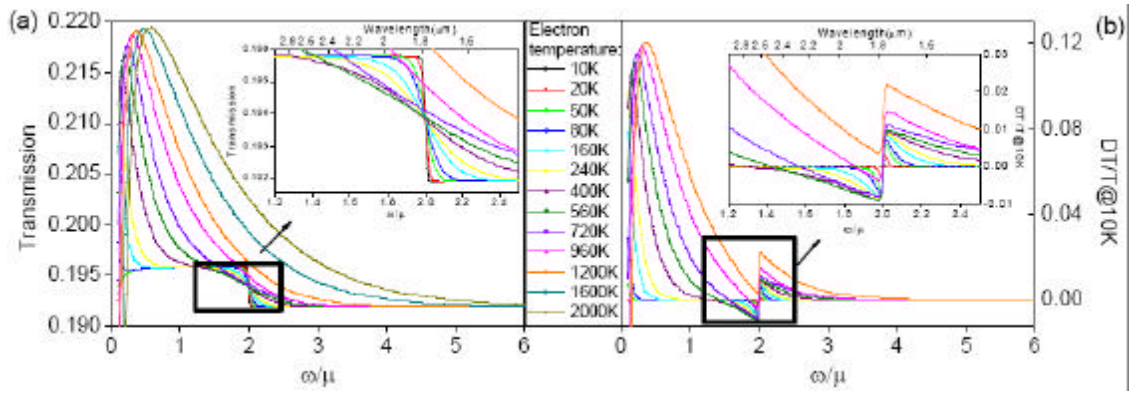

Fig. 4